# RAN Tester UE: An Automated Declarative UE Centric Security Testing Platform

[Dataset/Tool Paper]

Charles Marion Ueltschey*
Electrical and Computer Engineering
Mississippi State University
Starkville, Mississippi, USA
cmu32@msstate.edu

Joshua Moore*
Electrical and Computer Engineering
Mississippi State University
Starkville, Mississippi, USA
jjm702@msstate.edu

Aly Sabri Abdalla*
Electrical and Computer Engineering
Mississippi State University
Starkville, Mississippi, USA
asa298@msstate.edu

Vuk Marojevic*
Electrical and Computer Engineering
Mississippi State University
Starkville, Mississippi, USA
vuk.marojevic@msstate.edu

## Abstract

Cellular networks require strict security procedures and measures across various network components, from core to radio access network (RAN) and end-user devices. As networks become increasingly complex and interconnected, as in O-RAN deployments, they are exposed to a numerous security threats. Therefore, ensuring robust security is critical for O-RAN to protect network integrity and safeguard user data. This requires rigorous testing methodologies to mitigate threats. This paper introduces an automated, adaptive, and scalable user equipment (UE) based RAN security testing framework designed to address the shortcomings of existing RAN testing solutions. Experimental results on a 5G software radio testbed built with commercial off-the-shelf hardware and open source software validate the efficiency and reproducibility of sample security test procedures developed on the RAN Tester UE framework.

## CCS Concepts

• **Security and privacy** → **Systems security**; **Logic and verification**; **Penetration testing**; **Vulnerability scanners**; • **Software and its engineering** → **Software notations and tools**; **Software libraries and repositories**; • **Symposium on Access Control Models and Technologies**;

## Keywords

Automation, Integration, O-RAN, Testbed, Security, Software Radio.

## 1 Introduction

The rapid evolution of Radio Access Networks (RANs) toward open, disaggregated, softwarized, programmable, and artificial intelligence (AI)-driven frameworks introduces new security vulnerabilities. Unlike traditional monolithic RANs, O-RAN's open interfaces and software-defined components expose it to new threats, including protocol exploits, supply chain risks, and adversarial AI manipulation [1]. The integration of machine learning (ML) for use cases such as dynamic resource allocation and anomaly detection further complicates security, necessitating rigorous testing methodologies that can adapt to evolving threats [2]. Continuous and automated security testing can help protect network integrity, safeguard user data, and maintain reliable service availability.

The current security testing approaches often rely on static rule-based systems and manual penetration testing, which hinder their repeatability and consistency. Security testing for next-generation RANs must be reproducible, automated, and modifiable, enabling continuous assessment in diverse deployment scenarios. Existing works on automated security testing have focused on specific security tests, such as fuzzing, where a malicious, software-defined user equipment (UE) injects malformed packets to attack 5G O-RAN. This automated fuzzer is effective in disrupting the Radio Resource Control (RRC) setup [3]. The authors of [4] propose security testing methods to evaluate the resilience of the near-real time RAN Intelligent Controller (RIC) and the A1 interface, identifying potential security weaknesses via a customized A1 interface testing tool. However, these tests focus on a specific RIC deployment and are not automated. The work presented in [5] shows a non-automated transformer-based ML framework for assessing the impact of spoofing and replay attacks against precision time protocol synchronization in the O-RAN fronthaul for real-world and digital twin environments. Bonati, et al. [6] present an initial step toward automating the deployment and testing of a cloud-native, open, programmable, and multi-vendor cellular network. However, the framework does not consider security testing to assess the performance of next-generation RAN security measures.

ORANalyst [7], a security testing framework, systematically analyzes the robustness of O-RAN implementations by leveraging dynamic tracing, static analysis, and grammar-aware fuzzing to identify vulnerabilities, including crashes and messaging disruptions. However, its testing approach remains stateless, limiting its ability to analyze long-term interactions. While it automates input generation and dependency analysis, it still requires manual validation of findings, preventing full automation. 5G-CT [6] is an automated framework for end-to-end testing of softwarized 5G and O-RAN systems. It leverages Red Hat OpenShift and GitOps to streamline deployment and validation, which enables long-term automated conformance testing of open-source protocol stacks. The 5G security testing framework [8] focuses on testing the UE as opposed to testing the RAN, which is the goal of our research. Automated UE testing has been shown in another work that fully automates UE security testing [9].

Unlike previous works, we introduce a comprehensive security testing platform that adopts a UE-centric approach, enabling a more realistic assessment of real-world attack vectors in white-box and black-box RAN deployments. Our approach integrates an automated testing framework that streamlines security evaluations and enhances reproducibility. We provide a detailed architectural breakdown of the platform, outlining its key components and their roles in generating, detecting, and mitigating security threats. Furthermore, we demonstrate the platform's effectiveness through experimental deployment, comparing its performance and security assessment capabilities against a standard software radio setup. This comparison highlights the advantages of our system in providing deeper insights into security vulnerabilities while maintaining efficiency and scalability of testing procedures.

The remainder of this paper is structured as follows: Section II defines the key requirements for evaluating O-RAN security testing platforms and provides a detailed overview of the RAN Tester UE (RT UE) architecture and its core components. Section III introduces the platform's automated testing capabilities, highlighting its declarative environments and process management. Section IV presents the experimental deployment and discusses the platform's performance and security results. Section V offers concluding remarks and outlines potential future research. Our code for the RAN Tester UE framework is available on GitHub[1].

## 2 Requirements and Platform Architecture

### 2.1 Requirements for Reproducible and Automated RAN Security Testing

Security testing of RAN deployments must be designed to be reproducible, automated, and adaptable to evolving threats. Table 1 outlines the key requirements necessary for an effective security testing framework, ensuring robust and scalable assessments of network vulnerabilities.

- **Reproducibility:** System must reliably reproduce complex attack scenarios, ensuring consistent and valid test outcomes.

---

[1]https://github.com/oran-testing/ran-tester-ue

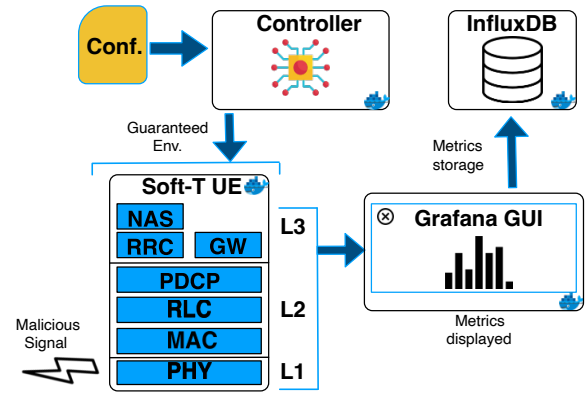

Figure 1: RAN Tester UE architecture.

- **Automation:** Automate large-scale security evaluations with minimal human intervention.
- **Modularity:** A modular security testing framework with standardized application programming interfaces (APIs) allows seamless integration of new test cases and attack vectors without altering core architecture.
- **Real-Time Analysis:** Real-time monitoring and analysis provide immediate feedback on security threats to RAN components.
- **Scalability:** Practical security testing tools need to be able to accommodate various deployment scenarios, from small-scale testbeds to full-scale production networks.
- **Standard Compliance:** Security testing must align with industry guidelines and specifications set by relevant organizations, such as 3GPP and the O-RAN Alliance.
- **Security Containment**: Isolation mechanisms, such as sandboxing and network segmentation, safeguard test integrity by preventing interference with external systems.

### 2.2 Architecture of RAN Tester UE

The RT UE platform builds upon the Soft T-UE security testing tool [11] designed for evaluating vulnerabilities in O-RAN environments. While Soft-T UE focuses on specific security tests, this work extends its capabilities into a modular and scalable security testing platform. It integrates Soft T-UE's tests while introducing broader security evaluation, automation, and real-time threat analysis for O-RAN deployments. These extensions offer a more adaptable and extensible solution for rigorous, reproducible security testing.

- **Controller:** The controller translates declarative configurations into guaranteed environments for launching attacks. It starts necessary components, including UEs for impact measurement, and configures them within Docker containers, passing required radio devices and network adapters. This ensures reproducibility of complex attacks and simplifies configuration.
- **Declarative Configurations:** YAML configurations are used for the controller. This allows for a simple hierarchical organization and alignment with the structure of the configuration data. Each process in the specified in the controller configuration can have customized configuration fields, allowing for a complete description of the required environment.
- **Real-time Data and Analysis:** The system incorporates InfluxDB and Grafana to allow for efficient real-time metrics and

Table 1: Key requirements for reproducible and automated security testing.

| Requirement | Description | Metrics |
| --- | --- | --- |
| Reproducibility | Ensures consistent security assessments across different test setups, enabling verification and comparison of vulnerabilities [10]. Guaranteed environments ensure that complex attacks with many components work well on any machine. Configuration files enable reproducing security tests. | Ability to reproduce an attack with a given configuration file; consistency in attack execution across different testbeds. |
| Automation | Reduces manual effort by enabling continuous and large-scale security evaluations with minimal human intervention. Facilitates the running of complex attacks with many components without delay or human error [6] so that every attack or test is run in the same way and with the same delay across systems. | Percentage of tests executed without manual intervention; time required to set up and execute an attack. |
| Modularity | Supports integration of new security tests and adaptability to emerging attack vectors. A plug-and-play API allows developers of all skill levels to easily integrate new components into the system. Standardized worker thread interfaces facilitate ease of integration for any component. | Number of new attack modules integrated without modifying the core system; ease of API use measured by lines of code required for integration. |
| Real-Time Analysis | Provides immediate feedback on security threats, enabling dynamic countermeasures. Gives insight into channel quality and the effects of tests on individual UEs. Allows the user to have a full view of the impact of various attacks on the RAN. | Latency in detecting and reporting attack impact; real-time monitoring accuracy. |
| Scalability | Allows testing across various deployment scenarios, from small-scale lab environments to full-scale production networks. The efficient and automated nature of the system allows for a huge number of tests with many components to be run on the same system. | Maximum number of concurrent tests supported; resource utilization efficiency across multiple test instances. |
| Standard Compliance | Aligns with security guidelines and specifications set by organizations such as 3GPP and the O-RAN Alliance. | Adherence to 3GPP and O-RAN security standards [10]; compatibility with standardized security frameworks. |
| Security | Ensures that attacks remain contained within the test environment and do not unintentionally disrupt external systems. Implements isolation mechanisms to prevent unintended network interference and safeguards test integrity. | Degree of attack containment; effectiveness of sandboxing and network isolation techniques. |

logs collection. Metrics handling is IP based and allows for a unique connection between each system component and the centralized metrics database. This approach increases resiliency by eliminating the single point of failure in traditional metrics collection mechanisms.

- **Metrics Database:** InfluxDB is selected for its efficient handling of time-series data and flexible NoSQL management. This enables trend analysis using the flux query language, ideal for security testing where data variety does not fit relational databases.
- **Metrics Presentation:** Grafana is used for real-time metrics display because of its versatile user interface, allowing for easy changes and quick development of various dashboards for each attack component. This facilitates a wide variety of visual representations.
- **Uu Interface Agent:** This is a lightweight in-phase/quadrature (IQ) data collector that has been developed for spectrum utilization analysis. This component collects IQ in bursts to be converted into constellations and spectrograms.
- **Attacks:** The RT UE system incorporates several attack components, including a customized UE based on srsRAN, a 5G Physical Downlink Control Channel (PDCCH) sniffer, and a jammer. The Soft T-UE based UE enables fuzzing attacks to the RRC and denial of service (DoS) attacks to the Random Access Channel (RACH) procedures. It implements the functions of a legitimate UE and executes targeted attacks during the setup process. The system also includes a resource-efficient 5G PDCCH sniffer, adapted from Ludant et al. [12], which gathers Downlink Control Information (DCI) to inform attacks using other components, such as providing downlink synchronization for targeting black and white box RANs.

## 3 Automated Testing

### 3.1 Declarative Environment

The system employs declarative configurations to ensure reproducible attacks. These configurations define both the required environment for each component and the execution sequence, guaranteeing correct dependency management. Each attack component can specify dependencies, ensuring that all required components are initialized beforehand. This approach eliminates execution order issues for attacks that rely on other system components.

Furthermore, the controller configuration explicitly defines paths and dependencies for each component. The controller automatically provisions the necessary docker images and interfaces according to the system configuration, preventing versioning conflicts and path-related issues common in manual deployments as well as facilitating scalability.

By adopting a declarative approach, RT UE minimizes the need for user-driven imperative management, reducing inconsistencies that arise from manual system configurations. Testing demonstrated that this approach consistently enables functional tests across machines of varying architectures and resource capacities, enabling reproducibility. In contrast, imperative testing requires additional debugging in the majority of cases. Even experienced operators face significant delays—often hours—when executing multi-component tests due to the complexity and instability of many wireless testing tools. RT UE, however, enables test execution with a single command, ensuring 100% automation across different environments, including MacOS and most linux distributions, without requiring additional troubleshooting or configuration.

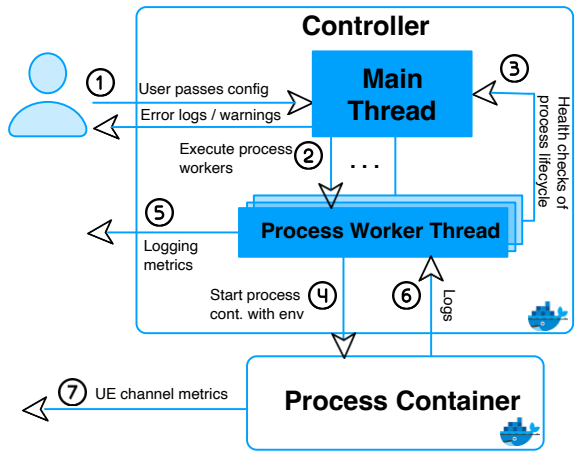

Figure 2: RAN Tester UE controller architecture.

### 3.2 Custom Thread-Based Process Lifecycle

A key feature that enables modularity, scalability, and security when running complex attacks is the controller's worker thread mechanism. When a security test is initiated, a dedicated thread is created for each attack component, allowing continuous monitoring and ensuring proper operation. The isolated nature of these worker threads improves modularity and security. Furthermore, this design enables RT UE to scale by executing any number of concurrent attacks while preserving the ability to restart or reconfigure individual components as needed. As illustrated in Fig. 2, in Step 1 the user passes the custom configuration to the main controller thread detailing the testing scenario. In Step 2, the main thread passes the necessary configuration to reproduce the expected environment. The worker thread then periodically signals the controller with the status of its assigned attack component in Step 3, allowing the main thread to notify the user when the system is incorrectly configured. In Step 4 the process container is started and begins to execute. A custom logging mechanism is employed in Step 5, providing real-time updates on the process lifecycle to maintain transparency and traceability. These logs are continuously synchronized in Step 6, as the process container relays information back to the corresponding worker thread. Finally, in Step 7, UE channel metrics are visualized in the Grafana GUI, allowing users to effectively analyze performance data and test outcomes. This structured approach ensures transparent error handling, robust zero-downtime deployments, and an intuitive user experience.

## 4 Experimental Deployment and Results

For our experimental analysis, we deploy the srsRAN Project gNodeB (gNB) in the split 7.2 O-RAN configuration on an AMD Ryzen 9 7950X, a 16-core processor with 100 GB of RAM, running Ubuntu 22.04 LTS with a real-time kernel. This machine hosts the central and distributed units (CU and DU), while a Falcon-RX switch manages the fronthaul, connecting to a Benetel RAN 550 radio unit (RU). The The RT UE software is deployed on a software radio platform featuring a host computer and an Ettus B210 USRP. It executes alongside a legitimate UE software stack, both containerized and interfacing with their respective B210 USRP for wireless transmission and reception. As presented in Fig. 3, this setup ensures a realistic testing environment, allowing for a comprehensive evaluation of the platform's capabilities. While developing and testing the platform, we are able to connect and test up to five software radio UEs with B210 USRPs running Soft T-UE and several commercial off-the-shelf UEs, functioning as a control group. Additionally, several USRPs are used for IQ data collection, as well as for implementing jamming and sniffing attacks.

### 4.1 Platform Performance

Figure 4 illustrates the enhanced efficiency, scalability, and usability of the RT UE platform across multiple metrics as compared to bare-metal deployments or existing security testing frameworks. Below, we summarize key performance improvements.

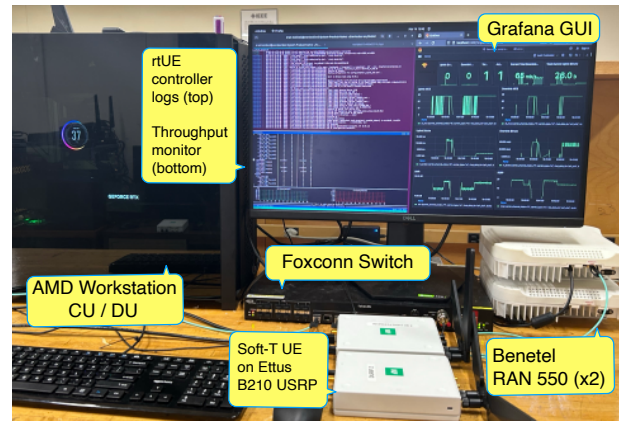

Figure 3: RAN Tester UE over-the-air testbed.

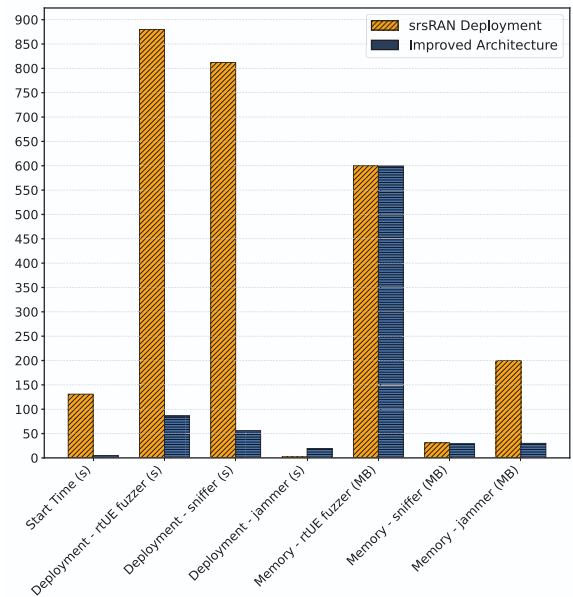

Figure 4: Performance metrics comparison between RT UE and srsRAN's UE.

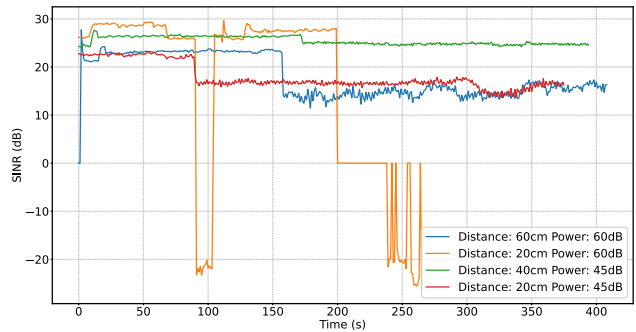

Figure 5: Collected SINR measurements of two UEs at various distances and different jamming attack powers.

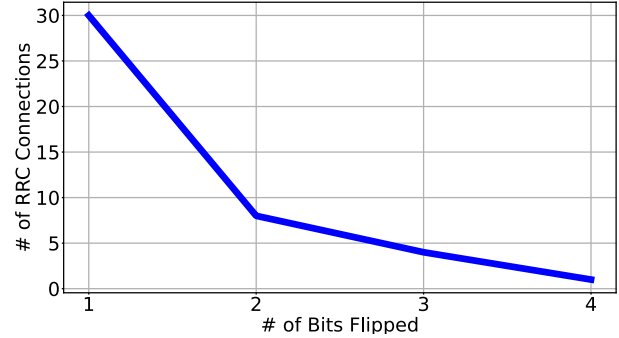

Figure 6: Number of successful RRC connections out of 100 attempts, when flipping 1 to 4 random bits.

- **Test Execution Time:** The new system completes tests in 5 s, compared to over 120 s in the previous version despite integrating additional attack components.
- **Deployment Time:** Deployment is reduced from over 28 minutes to 86 s, including the controller and metrics database. Concurrent deployments minimize overhead.
- **System Memory Utilization:** The RT UE maintains a stable 599 MB memory footprint, while optimized components, such as the jammer, reduce usage from 200 MB to 30 MB.
- **Disk Usage:** Traditional O-RAN security tests required multiple machines and 20 GB of storage. The new system consolidates operations to a single machine using only 7 GB.
- **Cross-Platform Support:** Fully containerized architecture removes Ubuntu 24.04 dependency, enabling compatibility with macOS and Windows.
- **Automation Capabilities:** The new system fully automates attack execution and provides an API for test integration. Experiment start times show a 100x speedup due to concurrent test management.
- **Configuration Complexity:** Attack scenarios with jammers, sniffers, and malicious UEs previously required 164 lines of configuration, now reduced to 43. The hierarchical architecture ensures consistent parameter inheritance and interoperability.

### 4.2 Security Results

The improvements introduced by the RT UE architecture are also evident in the attacks we have developed thus far. The results of our tests are described below and presented in Fig. 5 and Fig. 6.

- **Jammer:** The RT UE system integrates a custom jammer optimized for UHD devices. Figure 5 illustrates the signal-to-interference-plus-noise ratio (SINR) of two UEs over time, collected via the metrics reporting system, where SINR degradation is observed until the UE is released from the RAN. Tests were conducted with jamming power set to 45 dB and 60 dB relative gain settings, with UEs positioned at varying distances to the jammer.
- **Fuzzer:** A modified Soft-T-UE, based on srsRAN's UE, is developed to execute targeted security attacks [11]. One such attack, RRC fuzzing, is tested within the RT UE framework by flipping bits in Service Data Unit (SDU) buffers of RRC Dedicated and Common Control Channel messages from the UE to the gNB. This technique disrupts gNB operations, exposing vulnerabilities susceptible to targeted exploits such as buffer overflows.
- **DCI Sniffer:** The sniffer, based on 5GSniffer [12], passively decodes PDCCH transmissions from a 5G base station to extract DCI and Radio Network Temporary Identifiers (RNTIs). This enables fine-grained analysis of resource scheduling, control signaling, and device identification. Testing has shown a 99% successful PDCCH decode rate, with three concurrent UEs achieving a 98% success rate. Extracted DCI data can be leveraged for real-time network monitoring and automated attacks requiring UE downlink synchronization.
- **RACH Flooding:** RT UE also implements a RACH flooding attack, where valid preambles with randomized indexes are transmitted to overwhelm the gNB's contention resolution process. Empirical results show that the gNB crashes in 5% of test cases, while in other instances, it remains operational for already connected UEs but blocks new registrations.

## 5 Conclusions and future work

This paper has introduced RT UE, a UE-centric security testing platform for evaluating O-RAN security vulnerabilities with improved realism and automation. Our system enhances existing approaches by integrating automated testing, reducing deployment complexity, and supporting cross-platform execution. Through O-RAN testbed deployment and rigorous testing, we have demonstrated significant improvements in test execution time, resource efficiency, and scalability compared to a standard srsRAN-based setup. These advancements make our platform a valuable tool for reproducible and large-scale security assessment of next-generation RANs. Future work will focus on integrating AI-driven orchestration to optimize resource allocation and test execution dynamically. We will expand the framework to address novel threat models and advanced adversarial techniques for emerging O-RAN vulnerabilities.

## Acknowledgments

This material is based upon work supported by the National Telecommunications and Information Administration (NTIA) under Award No. 28-60-IF012 and by the Office of Naval Research under Award No. N00014-23-1-2808.